\documentstyle[amssymb,aps,12pt]{revtex}

\begin{document}
\draft
\author{O. B. Zaslavskii}
\address{Department of Mechanics and Mathematics, Kharkov V.N. Karazin's National\\
University, Svoboda\\
Sq.4, Kharkov 61077, Ukraine\\
E-mail: aptm@kharkov.ua}
\title{Regular self-consistent geometries with infinite quantum backreaction in 2D
dilaton gravity and black hole thermodynamics: unfamiliar features of
familiar models }
\maketitle

\begin{abstract}
We analyze the rather unusual properties of some exact solutions in 2D
dilaton gravity for which infinite quantum stresses on the Killing horizon
can be compatible with regularity of the geometry. In particular, the
Boulware state can support a regular horizon. We show that such solutions
are contained in some well-known exactly solvable models (for example, RST).
Formally, they appear to account for an additional coefficient $B$ in the
solutions (for the same Lagrangian which contains also ''traditional''
solutions) that gives rise to the deviation of temperature $T$ from its
Hawking value $T_{H}$. The Lorentzian geometry, which is a self-consistent
solution of the semiclassical field equations, in such models, is smooth
even at $B\neq 0$ and there is no need to put $B=0$ ($T=T_{H}$) to smooth it
out$.$ We show how the presence of $B\neq 0$ affects the structure of
spacetime. In contrast to ''usual''\ black holes, full fledged thermodynamic
interpretation, including definite value of entropy, can be ascribed (for a
rather wide class of models) to extremal horizons, not to nonextreme ones.
We find also new exact solutions for ''usual'' black holes (with $T=T_{H}$).
The properties under discussion arise in the {\it weak}-coupling regime of
the effective constant of dilaton-gravity interaction. Extension of
features, traced in 2D models, to 4D dilaton gravity leads, for some special
models, to exceptional nonextreme black holes having no own thermal
properties.
\end{abstract}



\section{Introduction}

Black hole thermodynamics and physics of black holes with a scalar field are
among favorite research areas of Prof. J. D. Bekenstein, in which seminal
results \cite{bekenst}, \cite{beka} - \cite{bekc} belong to him. In the
present paper I try to combine both these lines in a quite unusual context.
I argue that, if quantum backreaction is essential but the semiclassical
approximation is still valid, in dilaton gravity there exist exceptional
situations in which a nonextreme black hole may have a temperature not
coinciding with the Hawking value. On the other hand, thermodynamic
properties of extreme black holes, found earlier only within the tree-level
approximation, can be justified on the one-loop level. In so doing, instead
of taking a given classical solution with finding subsequent small
corrections, we are pursuing the goal to find and analyze self-consistent
solutions of quantum backreaction equations.

As in four dimensions (4D) the full problem is very complicated, we exploit
in the most part of the present paper two-dimensional (2D) dilaton gravity
since the essence of matter becomes much more transparent within its
framework. In the absence of a full theory of quantum gravity such theories
have assumed especial significance. It turned out that they possess profound
main features inherent also to the 4D world. In particular, they contain
black hole solutions and describe their formation and evaporation due to the
Hawking effect \cite{callan}. This was one of the main reasons why 2D
dilaton models became so popular during last decade (for a recent reviews,
see, e.g. \cite{od}, \cite{dv}). Within the framework of such theories, one
can take into account one-loop effects in a self-consistent way and analyze
them directly in terms of differential equations derived from the action
principle. Moreover, some families of 2D theories are exactly integrable,
providing us with a remarkable tool for visualizing subtle effects of black
hole physics. Thus, using relatively simple exactly solvable 2D models
enabled us, without unnecessary mathematical complexity, to gain further
insight in the known phenomena, relevant for more realistic four-dimensional
physics\footnote{%
In this article we are dealing with semiclassical dilaton gravity with
account for backreaction of conformal fields and do not consider additional
scalar \cite{fil1}, \cite{fil2} \cite{eliz2}, Yang-Mills or fermion fields 
\cite{strobl}, \cite{pelzer}, theories nonlinear with respect to curvature 
\cite{eliz4}, etc., where exact integrability is achieved for the classical
case only.}.

Moreover, the simplicity of the models under discussion sometimes helps us
to find some qualitatively new features which were completely overlooked in
four-dimensional gravity. In particular, in the previous articles \cite
{nonext}, \cite{ext} we pointed out that there exist examples with infinite
stresses, developed by quantum backreaction on the Killing horizon,
consistent with the regularity of geometry in the vicinity of a horizon.
This feature does not have counterparts in general relativity (but may have
them, in principle, in 4D dilaton theory) and looks so unusual that deserves
further study. In the present paper we extend and enlarge on observations
made in \cite{nonext}, \cite{ext} and put them on a more firm basis. We
would like to stress that we do not invent some particular artificial models
to get exotic behavior, but, rather, more attentively analyze properties of
already known ones, which did not receive proper attention before. We
consider quite ''normal'' string-inspired Lagrangians, such as the
Russo-Susskind-Thorlacius (RST) one \cite{rst}. The solutions under
discussion contain one more parameter $B$ (as compared to the ''usual''
black hole solutions in the RST\ model) and in the particular case $B=0$ the
previously known solutions are recovered. A more general exactly solvable
model, that includes the RST one as a particular case, was considered by
Cruz and Navarro-Salas (CN) \cite{cruz}. We want to stress that the quantity 
$B$ is the parameter of a solution itself and does not appear in the
Lagrangian. Thus, actually, what is found in Refs. \cite{nonext}, \cite{ext}
is the property, intrinsically inherent to some popular models, which was
not paid attention to before.

That some divergencies of quantum stresses may occur in spite of regularity
of {\it self-consistent} solutions in 2D dilaton gravity, was already
pointed out in literature \cite{triv}. The corresponding divergencies are
rather weak in that they happen in the frame of a free-falling observer (not
in the original Schwarzchild-like one) and are related to extreme horizons
only. Meanwhile, the divergencies under consideration are much more severe
in the sense that they appear in the Schwarzchild-like frame and for the
nonextreme case as well. In spite of it, under some circumstances, they do
not spoil the regularity of the geometry on a horizon.

As a matter of fact, there exist works in which some concrete properties of
the aforementioned models (for instance, reaction of black holes to shock
waves \cite{flux}) were analyzed without, however, paying attention to the
rather curious relationship between regularity of the geometry and behavior
of quantum stresses in some classes of solutions. Meanwhile, this
non-trivial relationship, contradicting habitual expectations, deserves in
itself, in our view, separate discussion. It turns out that for some classes
of solutions the quantum stress-energy tensor at infinity $T_{\mu }^{\nu
(q)}\rightarrow \frac{\pi }{6}T^{2}diag(-1,1)$, $T\neq \frac{\kappa }{2\pi }$
($\kappa \,$is a surface gravity) without destroying a regular geometry near
the horizon. For black hole physics, it means extension of the types of
basic states (Hartle-Hawking, Unruh and Boulware ones) and possible
rearrangement of their properties in some new combinations. Say, regularity
of the geometry at the horizon (feature, inherent to the Hartle-Hawking
state) proves to be consistent with vacuum-like behavior of quantum stresses
at infinity, typical of the Boulware state (see Sec. IV\ C below). From the
thermodynamic viewpoint, the solutions under discussion represent an
exceptional case when the intimate connection between the surface gravity
and geometry is broken: usually, the unique choice of the temperature (for
nonextremal horizons) enables one to smooth out the geometry but now the
geometry is already smooth from the very beginning. Moreover, the attempt to
calculate the Euclidean action for the nonextreme case shows that for the
aforementioned models it is infinite and, thus, the temperature remains only
as a formal parameter, determining, say, the form of quantum stress-energy
tensor at infinity. Thus, we get Killing horizons without full fledged
thermodynamics. I restrict myself by static geometries but the existence of
the solutions with such properties poses the question about alternative
scenarios of black hole evaporation.

On the other hand, for extremal horizons of the type under consideration,
our configuration, on the contrary, seems to be the only possible case to
get a more or less reasonable thermodynamics and justify on the {\it %
semiclassical} level the prescription $S=0$ for the entropy, made earlier
for the {\it classical} case \cite{ross} - \cite{gib}: infinite stresses for
nonzero temperature are inevitable on extremal horizons and it is these
solutions which successfully ''cope with '' them. In other words, if for
''usual'' black holes thermodynamics is well-defined in the nonextreme case
and questionable in the extreme one, now the situation is completely
opposite.

What is said above can serve as motivation to look at the solutions at hand
without prejudice. They appear as an inevitable consequence of some 2D
models, viewed as closed systems, and are worth studying with all possible
completeness.

One reservation is in order. We discuss rather large number of different
model cases, but all this multiformity stems from the same root and reveals
the fact that, within the same model (mainly, the CN one), different
relationships between parameters (including degenerate cases, when some
parameters are taken to be zero) gives rise to qualitatively different
physical situations.

The paper is organized as follows. In Sec. II we write down basic field
equations of gravitation-dilaton system and discuss the structure of field
equation in the pure classical and semiclassical cases.

In Sec. III we summarize briefly the main features of the approach to
exactly solvable models of two-dimensional (2D) dilaton gravity with
backreaction. In so doing, we fill the gap, left in our previous papers \cite
{exact}, \cite{thr} and show that the conditions of exact solvability are
conformally invariant.

In Sec. IV we trace in detail, using the CN model as an example, how the
inclusion of the parameter $B$ changes the structure of spacetime and leads
to unbounded stresses on the horizon.

In Sec. V we suggest explicitly the model that admits solutions with
extremal horizons, possessing the properties under discussion. The
curvature-coupling function represents a combination of exponent of a
dilatonic field and, in this respect, can be in principle achieved in string
theory. Further, we consider the thermodynamics of {\it quantum-corrected }%
self-consistent extremal horizons, not restricting ourselves by the
particular model, and show that for some class of the models, the Euclidean
action for corresponding solutions is finite in spite of divergencies in
quantum stresses, and the entropy $S=0$.

In Sec. VI we discuss briefly the relevance of the issues under discussion
for more realistic 4D gravity.

In Sec. VII we summarize the main features of the solutions considered in
the paper.

\section{Basic equations and relationship between classical and
semiclassical quantities.}

Consider the gravitation-dilaton 2D theory taking into account effects of
backreaction of quantum massless fields. Then the bulk part of the total
action reads 
\begin{equation}
I_{V}=I_{0}(g_{\mu \nu };F,V,U)+I_{PL}(g_{\mu \nu },\psi )\text{,}
\label{tot}
\end{equation}
where 
\begin{equation}
I_{0}(g_{\mu \nu };F,V,U)=\frac{1}{2\pi }\int d^{2}x\sqrt{g}[RF(\phi
)+V(\phi )(\nabla \phi )^{2}+U(\phi )]\text{,}  \label{a}
\end{equation}
$R$ is a Riemann curvature. Quantum backreaction is described by the
Polyakov-Liouville action

\begin{equation}
I_{PL}(g_{\mu \nu },\psi )=-\frac{\kappa }{2\pi }\int_{M}d^{2}x\sqrt{-g}[%
\frac{(\nabla \psi )^{2}}{2}+\psi R]\text{,}  \label{pl}
\end{equation}
where $\kappa =\frac{\hslash N}{24}$, $N$ is number of quantum fields. It is
implied that $N\rightarrow \infty $, $
\rlap{\protect\rule[1.1ex]{.325em}{.1ex}}h%
\rightarrow 0$ in such a way, that $\kappa \,$is kept fixed. Due to large $N$
expansion, the contribution of higher loops and manifestation of quantum
properties of the dilaton field is suppressed, and the problem is reduced to
the analysis of a closed set of semiclassical equations which follow from
the action (\ref{tot}).

The equation for the auxiliary field $\psi $ that follows from (\ref{pl})
has the form 
\begin{equation}
\Box \psi =R\text{ .}  \label{psai}
\end{equation}

The field equations read $T_{\mu \nu }\equiv 2\frac{\delta I}{\delta g^{\mu
\nu }}=0$. The tensors corresponding to the parts $I_{0}\,$and $I_{PL}$of
the action (\ref{pl}) are equal to 
\begin{equation}
T_{\mu \nu }^{(0)}=\frac{1}{2\pi }[2(g_{\mu \nu }\square F-\nabla _{\mu
}\nabla _{\nu }F)-Ug_{\mu \nu }+2V\nabla _{\mu }\phi \nabla _{\nu }\phi
-g_{\mu \nu }V(\nabla \phi )^{2}]  \label{to}
\end{equation}
\begin{equation}
T_{\mu \nu }^{(PL)}=-\frac{\kappa }{2\pi }\{\partial _{\mu }\psi \partial
_{\nu }\psi -2\nabla _{\mu }\nabla _{\nu }\psi +g_{\mu \nu }[2\nabla
^{2}\psi -\frac{1}{2}(\nabla \psi )^{2}]\}\text{,}  \label{tpl}
\end{equation}
$T_{\mu \nu }=T_{\mu \nu }^{(0)}+T_{\mu \nu }^{(PL)}$. The dilaton equation
which is obtained by varying $\phi $, reads 
\begin{equation}
F^{\prime }R+U^{\prime }-2V\square \phi -V^{\prime }\left( \nabla \phi
\right) ^{2}=0\text{,}  \label{dil}
\end{equation}
where prime denotes differentiation with respect to $\phi $.

It is seen from (\ref{tpl}) that $T_{\mu }^{\mu (PL)}=-\frac{\kappa }{\pi }R$%
. In the static situation both $\phi $ and $\psi $ depend on a spatial
coordinate only. As a result, the semiclassical action and field equations
retain the classical form but with the shifted coefficients: 
\begin{equation}
V\rightarrow \tilde{V}=V-\frac{\kappa \psi ^{\prime 2}}{2},F\rightarrow 
\tilde{F}=F-\kappa \psi  \label{ren}
\end{equation}
and $T_{\mu \nu }(F,V,U)=T_{\mu \nu }^{(0)}(\tilde{F},\tilde{V},\tilde{U})$, 
$\tilde{U}\equiv U$. In a similar way, the dilaton equation (\ref{dil})
retains its form in terms of tilded quantities: 
\begin{equation}
\tilde{F}^{\prime }R+U^{\prime }-2\tilde{V}\square \phi -\tilde{V}^{\prime
}\left( \nabla \phi \right) ^{2}=0.  \label{dilt}
\end{equation}

If, for example, we use the Schwarzschild gauge, 
\begin{equation}
ds^{2}=-dt^{2}f+f^{-1}dx^{2}\text{,}  \label{sc}
\end{equation}
we get for the static metric the $00$ and $11$ equations read 
\begin{equation}
2f\frac{\partial ^{2}\tilde{F}}{\partial x^{2}}+\frac{\partial f}{\partial x}%
\frac{\partial \tilde{F}}{\partial x}-U-\tilde{V}f\left( \frac{\partial \phi 
}{\partial x}\right) ^{2}=0\text{,}  \label{sc00}
\end{equation}

\begin{equation}
\frac{\partial f}{\partial x}\frac{\partial \tilde{F}}{\partial x}-U+\tilde{V%
}f\left( \frac{\partial \phi }{\partial x}\right) ^{2}=0\text{.}
\label{sc11}
\end{equation}

Thus, the system (\ref{dilt}), (\ref{sc00}), (\ref{sc11}) has the same form
as in the classical case, but with actions coefficients replaced by their
tilded counterparts.

The quantity $\psi $ can be found from (\ref{psai}): 
\begin{equation}
\frac{\partial \psi }{\partial x}=\frac{b-\frac{\partial f}{\partial x}}{f}%
\text{,}  \label{psx}
\end{equation}
The stress-energy tensor

\begin{equation}
T_{1}^{1(PL)}=-\frac{\pi }{6f}[T^{2}-(\frac{f_{x}^{\prime }}{4\pi })^{2}]%
\text{.}  \label{t1}
\end{equation}
For a black hole spacetime its behavior is intimately connected with the
properties of the quantum state \cite{solod}. For the Hartle Hawking state $%
b=\left( \frac{\partial f}{\partial x}\right) _{H}=4\pi T_{H}$, index ''H''
refers to the horizon, where $\psi $ and its derivative remain bounded, and $%
T_{H}$ is the Hawking temperature.

\section{Exact solvability and conformal properties of the action}

A special role in 2D dilaton gravity is played by models which are exactly
solvable semiclassically, with quantum backreaction taken into account.
Usually, such models represent some ''deformation'' of the classical CGHS
Lagrangian \cite{callan} due to terms containing $\kappa $ explicitly. In
view of importance of exactly solvable models, hereafter we mainly
concentrate just on such models.

According to \cite{exact}, \cite{thr}, the condition of exact solvability
can be written in the form 
\begin{equation}
D(u,\omega ,V)\equiv u^{\prime }(2V-\omega u)+u(u\omega ^{\prime }-V^{\prime
})+\kappa (\omega V^{\prime }-2V\omega ^{\prime })=0\text{,}  \label{d}
\end{equation}
where by definition 
\begin{equation}
u\equiv F^{\prime }(\phi ),U\equiv \Lambda \exp (\int d\phi \omega ).
\label{U}
\end{equation}
Equation (\ref{d}) can be solved: 
\begin{equation}
V=\omega (u-\frac{\kappa \omega }{2})\text{.}  \label{v}
\end{equation}
Then it turns out that 
\begin{equation}
\psi =\psi _{0}+\gamma \sigma (\phi )\text{, }\psi _{0}=\int d\phi \omega
(\phi )=\ln U\text{, }\gamma =const\text{.}  \label{ps}
\end{equation}
Here $\sigma $ has the meaning of a spatial coordinate in the conformal
gauge \cite{thr}: 
\begin{equation}
ds^{2}=f(-dt^{2}+d\sigma ^{2})\text{.}  \label{cmet}
\end{equation}

For the exactly solvable models found in \cite{exact} $\gamma =0$. For other
types of exact solutions $\gamma \neq 0$. There is no contradiction here
since the function $\psi $, obeying eq. (\ref{psai}), is ambiguous, being
defined up to a function $\sigma $ satisfying the relation $\square \sigma
=0 $. Different choices of the constant $\gamma $ correspond to the
different choices of the physical state of a system. Thus, for usual black
holes $\gamma =0$ and $\psi $ is finite on the horizon \cite{exact}, for
semi-infinite throats $\gamma \neq 0$, $\sigma \rightarrow -\infty $ and $%
\psi $ diverges there \cite{thr}; in a similar way these quantities behave
at the horizon of the ''singularity without singularities'' solutions in 
\cite{nonext}, \cite{ext}.

If one uses the conformal gauge, for the exactly solvable models under
discussion

\begin{equation}
\tilde{F}^{(0)}\equiv F-\kappa \psi _{0}=C+De^{-\sigma \delta }+\kappa
\gamma (1-\frac{\gamma }{2\delta })\sigma \text{,}  \label{F}
\end{equation}
\begin{equation}
f=e^{-\psi _{0}-\delta \sigma }\text{.}  \label{fd}
\end{equation}
Here $\delta $ is some constant. It is related to $\Lambda $ (\ref{U})
according to the relationship 
\begin{equation}
D\delta ^{2}=\Lambda  \label{const}
\end{equation}
that follows from 
\begin{equation}
U=\Box \tilde{F}=\Box \tilde{F}^{(0)}  \label{uf}
\end{equation}
that, in turn, follows from the field equations (see \cite{thr} for
details). It is convenient to introduce the dimensionless coordinate. Let,
for definiteness, 
\begin{equation}
\Lambda =4\lambda ^{2}>0\text{.}  \label{ul}
\end{equation}
Then one can achieve $D=1$ by a suitable shift in a coordinate and (\ref{F})
with $\delta =-2\lambda $ can be rewritten as 
\begin{equation}
\tilde{F}^{(0)}(\phi )=h(y)\equiv e^{2y}-By+C\text{, }B=-\kappa \frac{\gamma 
}{\lambda }(1+\frac{\gamma }{4\lambda })\text{, }y\equiv \lambda \sigma
\label{sol}
\end{equation}
\begin{equation}
f=\frac{e^{2y}}{U(\phi )}\text{.}  \label{gu}
\end{equation}

The curvature 
\begin{equation}
R=-f^{-1}\lambda ^{2}\frac{\partial ^{2}\ln f}{\partial y^{2}}\text{.}
\label{cr}
\end{equation}
If, at some point $\phi _{0}$, $\tilde{F}^{(0)\prime }(\phi _{0})=0$, the
scalar curvature $R$ diverges there, as it follows from (\ref{sol}) - (\ref
{cr}). Thus, $\phi _{0}$ is a singularity.

Usually, in investigations of the structure of dilaton-gravity theories an
important role is played by the conformal transformations, in the process of
which all three action coefficients $F$, $V$, $U$ change (see, e.g. \cite
{medkunst} and the literature quoted there). Therefore, the natural question
arises, whether the formulas (\ref{d}) and (\ref{ps}), (\ref{gu}) (the
latter two involve the coefficient $U$ only, but not $F$ and $V$) are
independent of the choice of a conformal frame. Below, we show that this is
indeed the case. To this end, let us consider the conformal transformation 
\begin{equation}
g_{\mu \nu }=e^{2\chi (\phi )}\bar{g}_{\mu \nu }\text{.}  \label{g}
\end{equation}
Then $\sqrt{g}=\sqrt{\bar{g}}e^{2\chi }$ and 
\begin{equation}
\sqrt{g}R=\sqrt{\bar{g}}(\bar{R}-2\bar{\nabla}^{2}\chi )\text{.}  \label{r}
\end{equation}
After substitution of (\ref{g}), (\ref{r}) into (\ref{a}), (\ref{pl}) and
integration by parts we obtain again an action of the form (\ref{tot}) 
\begin{equation}
I=I_{0}(\bar{g}_{\mu \nu };\bar{F},\bar{V},\bar{U})+I_{PL}(\bar{g}_{\mu \nu
},\bar{\psi})\text{,}
\end{equation}
with the redefined quantities: 
\begin{equation}
\bar{F}=F+2\kappa \chi \text{, }\bar{V}=V+2u\chi ^{\prime }+2\kappa \chi
^{\prime 2}\text{, }\bar{U}=Ue^{2\chi }\text{, }\bar{\psi}=\psi +2\chi (\phi
)\text{. }  \label{redef}
\end{equation}

The prime denotes differentiation with respect to $\phi $. This means that 
\begin{equation}
\bar{u}=u+2\kappa \chi ^{\prime },\bar{\omega}=\omega +2\chi ^{\prime }\text{%
.}  \label{u}
\end{equation}

Then it is straightforward to check that $D(\bar{u},\bar{\omega},\bar{V}%
)=D(u,\omega ,V)$. Therefore, the conditions (\ref{d}) and (\ref{ps}), (\ref
{gu}) are indeed conformally invariant.

\section{CN model}

For this model \cite{cruz} 
\begin{equation}
F=\exp (-2\phi )+2\kappa (d-1)\phi \text{, }V=4\exp (-2\phi )+2(1-2d)\kappa 
\text{, }U=4\lambda ^{2}\exp (-2\phi )\text{, }\omega =-2\text{.}  \label{cn}
\end{equation}
One can easily check that the condition (\ref{v}) is fulfilled for this
model. In the case $d=1/2$ it turns into the RST\ \cite{rst} model, for $d=0$
$\,$it becomes the BPP one \cite{bose}. In the conformal gauge we have,
according to (\ref{sol}), (\ref{gu}): 
\begin{equation}
\tilde{F}=\exp (-2\phi )+2\kappa d\phi =e^{2y}-By+C\equiv h(y)\text{,}
\label{eq}
\end{equation}
\begin{equation}
f=e^{2y+2\phi }  \label{cg}
\end{equation}
(the metric coefficient $f$ is defined up to the factor), the scalar
curvature is given by 
\begin{equation}
R=-2\lambda ^{2}f^{-1}\frac{d^{2}\phi }{dy^{2}}\text{.}  \label{R}
\end{equation}

Let us write $B\equiv b\kappa $. In fact, only the case $b=1$ was discussed
in \cite{cruz}, \cite{flux}. Meanwhile, we will keep $b$ as a free
parameter. The horizon, if it exists, lies at $y=-\infty $. It turns out
that the solutions of the type \cite{nonext} exist only for $d>0$, so we
restrict ourselves to this case.

In another popular coordinate set $\pm \lambda x_{\pm }=e^{\pm \sigma _{\pm
}}$, $\sigma _{\pm }=t\pm \sigma $, eq. (\ref{eq}) becomes 
\begin{equation}
\exp (-2\phi )+2\kappa d\phi =-\lambda ^{2}x_{+}x_{-}-\frac{B}{2}\ln
(-\lambda ^{2}x_{+}x_{-})+C\text{.}  \label{+-}
\end{equation}

\subsection{Spacetime structure for $B=0$}

First, if $b=0$, the dilaton field takes the finite value $\phi _{h}$ on the
horizon according to $\exp (-2\phi _{h})+2\kappa d\phi _{h}=C$. For the
metric one finds from (\ref{cg}) 
\begin{equation}
f=1-Ce^{2\phi }+2\kappa d\phi e^{2\phi }\text{.}  \label{gsol}
\end{equation}
It is convenient now to introduce, instead of the conformal coordinate $y$,
the Schwarzschild one $x=\lambda ^{-1}\int dyf$. Then the metric take the
form (\ref{sc}). In our case 
\begin{equation}
\lambda x=-\phi +\frac{\kappa d}{2}e^{2\phi }+const\text{.}
\end{equation}
Then there are two branches of the solution $\phi _{1}(x)$ and $\phi _{2}(x)$%
, glued along the singularity at $\phi =\phi _{0}$, $x=x_{0}$. The detailed
analysis was performed in \cite{solod} for the RST\ model, when $d=1/2$.
There is no qualitative difference here between the RST model and generic $%
d>0$, so we only repeat the main properties of the solutions briefly (see
also \cite{exact} for more general discussion).

1a. $\tilde{F}(\phi _{0})\equiv C_{0}>C$. The low branch: $\phi \in $ $%
(-\infty ,\phi _{0}]$, $x\in (\infty ,x_{0}).$ The upper one: $\phi \in $ $%
[\phi _{0},\infty )$, $x\in [x_{0},\infty )$, the point $x_{0}$ corresponds
to $\phi _{0}$, where the spacetime is singular. At the infinity we have,
for the lower branch, the linear dilaton vacuum $\phi =-y$, the spacetime is
Minkowskian. For the upper branch, the metric at infinity also
asymptotically approaches the flat spacetime but now $f\sim (l\ln l)^{2}$,
where $l=\int dy\sqrt{f}$ is a proper length. The horizons are absent, and
the singularity at $\phi _{0}$ is naked.

1b. $C_{0}<C$.

There are two regular horizons at $\phi _{1}$ and $\phi _{2},$ where $\tilde{%
F}(\phi _{1})=\tilde{F}(\phi _{2})=C$, $\phi _{1}<\phi _{0}<\phi _{2}$. For
each branch, the singularity is hidden behind the horizon. Both horizons
share the same Hawking temperature $T_{\lambda }=\lambda /2\pi $.

1c. $C_{0}=C$.

There exists only one singular horizon at $\phi =\phi _{0}$.

\subsection{Case $B\neq 0$}

It is instructive to trace what new features are brought about by
introducing $B\neq 0$. We assume that $B>0$ since it is this case that
corresponds to the solutions with regular geometries and infinite stresses
(see below).

Now the function $h(y)$ is not monotonic, as it was for $B=0$; it has a
minimum at $y_{0}=\frac{1}{2}\ln \frac{B}{2}$, $h(y_{0})=\frac{B}{2}(1-\ln 
\frac{B}{2})+C\equiv C_{1}$.

1a. $C_{0}>C_{1}$

There is the singular point at $\phi =\phi _{0}$, $y=y_{0}$, from which two
branches exit to $y\rightarrow \infty $ and two extend to $y\rightarrow
-\infty $ . To the right from this point the asymptotic behavior of both
branches at infinity does not change qualitatively as compared to the
property 1.a of the case $B=0$ since in $h(y)$ (\ref{sol}) it is the
exponent which dominates, whereas the term $By$ is negligible. To the left
of the singularity there is a singular horizon on the low branch at $%
y\rightarrow -\infty $, $\phi \rightarrow -\infty $: $f\sim l^{2}\rightarrow
0$ ($l$ is the proper distance from the singularity), $R$ $\sim -(l\ln
l)^{-2}\rightarrow -\infty $. As far as the upper branch is concerned, its
asymptotic nature (at $y\rightarrow -\infty $, $\phi \rightarrow \infty $)
depends on the value of $B$. Indeed, it follows from (\ref{eq}) - (\ref{R})
that for this branch 
\begin{equation}
f\sim e^{2y(1-\rho )},R\sim -e^{2y(-1+2\rho )}\text{, }\rho =\frac{B}{%
2\kappa d}\text{.}  \label{hor}
\end{equation}
Therefore, the horizon exists only for $\rho <1$. If $\frac{1}{2}<\rho <1$, $%
R\rightarrow 0$ and the geometry near the horizon is regular; if $\rho =%
\frac{1}{2}$, $R\rightarrow const<0$. For $\rho <\frac{1}{2}$ the curvature $%
R$ diverges and we have a singular horizon. Thus, a regular horizon exists
if 
\begin{equation}
\frac{1}{2}\leq \rho <1\text{.}  \label{cond}
\end{equation}

1b. $C_{0}<C_{1}$.

What is said in the property 1a about the behavior of the metric in
asymptotic regions of spacetime retains its validity. However, now the naked
singularity at $\phi =\phi _{0}$ is absent; we have two disjoint branches $%
\phi _{1}(y)$ and $\phi _{2}(y)$, each of which extends from $y=-\infty $ to 
$y=+\infty $.

1c. $C_{0}=C_{1}$.

This case is especially interesting. In the vicinity of $\phi _{0}$ both the
right and left hand sides of eq. (\ref{sol}) behave quadratically.
Therefore, there are two branches intersecting in the point $\phi _{0}$ with
different finite slopes: $\phi -\phi _{0}=\pm \frac{h^{\prime \prime }(y_{0})%
}{\tilde{F}^{(^{\prime \prime }0)}}$. As far as the behavior at $%
y\rightarrow \pm \infty $ is concerned, the analysis of the property 1a
applies. Thus, the point $\phi =\phi _{0}$ now becomes regular (let us
recall that in the case $B=0$ it was the point of the singular horizon, in
which two branch of the solution glued). There are two branches, one of
which extends from a singular horizon at $y=-\infty $, $\phi =-\infty $ to
the asymptotically flat region, where $f\sim (l\ln l)^{2}$. The second
branch corresponds to the Minkowski spacetime at $y=\infty $ and, dependent
of whether the condition (\ref{cond}) is fulfilled or not, it can possess
either a regular or singular horizon at $y=-\infty $. (If $\rho >1$, there
is no horizon at all at $y\rightarrow -\infty $. In this case configurations
like semi-infinite throats are possible \cite{bose}, \cite{thr} but we will
not discuss these cases here.)

It is convenient to summarize these observations in the table (recall that
for $B=0$ the quantity $C_{1}=C$):

$
\begin{tabular}{|l|l|l|}
\hline
& $B=0$ & $B>0$ \\ \hline
$C_{0}>C_{1}$ & $(NS,M),(NS,A)$ & $(NS,M),(NS,A),(SH,NS),[(H,NS)$ or $%
(SH,NS)]$ \\ \hline
$C_{0}<C_{1}$ & $(HS,A),(HS,M)$ & $[(SH,A)$ or $(H,A)],(SH,M)$ \\ \hline
$C_{0}=C_{1}$ & $(SH,M),(SH,A)$ & $[(H,M)$ or $(SH,M)]$, $(SH,A)$ \\ \hline
\end{tabular}
$

Here $NS$ means naked singularity, $SH\,$- singular horizon, $HS$ -
singularity, hidden behind the regular horizon, $H\,$- regular horizon, $M$
- asymptotically Minkowski region, $A$ - the region with the asymptotic
metric $f\sim \left( l\ln l\right) ^{2}$ at $l\rightarrow \infty $. For
example, $(H,NS)$ denotes the branch that extends from a regular horizon at $%
y\rightarrow -\infty $ to a naked singularity, and so on.

\subsection{Behavior of quantum stresses and mechanism of cancellation.
Regular horizons supported by quantum fields in Boulware state.}

We saw above that the behavior of the metric and dilaton changes
qualitatively in the vicinity of the horizon if $B\neq 0$. Indeed, for $B=0$
it follows from (\ref{eq}) that near the horizon, when $y\rightarrow -\infty 
$, $\phi $ tends to a finite value, while for $B\neq 0$ for the upper branch
in the main approximation $\phi =-\frac{B}{2\kappa d}y\rightarrow \infty $,
however small $B$ is$.$ If in (\ref{hor}) $B\rightarrow 0$, the metric
exhibit singular behavior ($R$ diverges), whereas if $B=0$ from the very
beginning, the horizon is regular. On the other hand, if $B\neq 0$, the
horizon is regular only provided $B$ is large enough: according to (\ref
{cond}), $B>\kappa d$.

It is also instructive to trace in more detail how the existence of regular
horizons with infinite quantum backreaction can follow from the structure of
field equations. Consider a generic action $I=I_{gd}+I_{m}$, where the
gravitation-dilaton part (not necessarily two-dimensional) has the same form
as (\ref{a}) and $I_{m}$ is the contribution of matter fields. Then one can
infer the field equations by varying the metric ($T_{\mu }^{\nu (m)}=2\frac{%
\delta I_{m}}{\delta g^{\mu \nu }}$): 
\begin{equation}
2FG_{\mu }^{\nu }+\theta _{\mu }^{\nu }=16\pi T_{\mu }^{\nu (m)}  \label{feq}
\end{equation}
\begin{equation}
\theta _{\mu }^{\nu }\equiv 2(\delta _{\mu }^{\nu }\square F-\nabla _{\mu
}\nabla ^{\nu }F)-U\delta _{\mu }^{\nu }+2V\nabla _{\mu }\phi \nabla ^{\nu
}\phi -\delta _{\mu }^{\nu }V(\nabla \phi )^{2}\text{.}  \label{feq1}
\end{equation}

In the case of general relativity $F=1$, $U=V=0=\theta _{\mu }^{\nu }$, so
the field equations take the form 
\begin{equation}
G_{\mu }^{\nu }=8\pi T_{\mu }^{\nu (m)}  \label{tgr}
\end{equation}
If $T\neq T_{H}$, the right hand side of (\ref{tgr}) diverges on the horizon
that is obviously incompatible with the regularity of $G_{\mu }^{\nu }$.
This is just an explanation of why one {\it must} put $T=T_{H}$.

In the case of 2D dilaton gravity the Einstein tensor $G_{\mu }^{\nu }\equiv
0$. If all the action coefficients $F$, $U$, $V$ are regular near the
horizon, so is $\theta _{\mu }^{\nu }$ and the proof retains its validity.
The only difference is that the divergencies of the quantum stress-energy
tensor, having the same magnitude as that for thermal radiation, go like $%
T_{loc}^{4}$ in the 4D case and like $T_{loc}^{2}$ in the 2D one, where $%
T_{loc}=T/\sqrt{-g_{00}}$ is the local Tolman temperature near the horizon.
However, if in (\ref{feq}), (\ref{feq1}) the quantities $F$, $\theta _{\mu
}^{\nu }$ themselves diverges near the horizon, the situation may change
drastically. Let these quantities have the asymptotics $\theta _{\mu }^{\nu
}\simeq t_{\mu }^{\nu }f^{-1}$, with some constants $t_{\mu }^{\nu }$. Take
into account that the Polyakov-Liouville tensor has the same asymptotics
(see below for details) $T_{\mu }^{\nu (PL)}\simeq t_{\mu }^{\nu (PL)}f^{-1}$%
. Then the set of field equations (\ref{feq}), (\ref{feq1}) now becomes 
\begin{equation}
\frac{t_{\mu }^{\nu }}{f}=\frac{t_{\mu }^{\nu (PL)}}{f}+b_{\mu }^{\nu }\text{%
, }  \label{comp}
\end{equation}
where $b_{\mu }^{\nu }$ is the part finite on the horizon (its concrete form
is now irrelevant). Multiplying (\ref{comp}) by $f$, we get that this
equation is self-consistent provided $t_{\mu }^{\nu (PL)}=t_{\mu }^{\nu }$.
In other words, the solutions under discussion are possible if the
contribution of the gravitation-dilaton part near the horizon into field
equations has {\it the same order }$f^{-1}$ as that of quantum fields. In
fact, explicit solutions of such a type were found (without analysis of
behavior of separate contributions in \cite{cruz}; some their particular
properties were discussed in \cite{flux}).

Actually, this imposes a restriction on the parameters of the model and may
or may not be fulfilled. If the solution does exist, it just means that we
have a metric regular near the horizon since (i) all quantities entering the
field equations were calculated with respect to a regular metric, and (ii)
this set of equations is self-consistent.

It is instructive to list now the explicit formulas for the stress-energy
tensor and link its properties with the relationship between the temperature
of quantum fields $T$ and the Hawking temperature $T_{H}$. Integrating (\ref
{psai}), one finds for static geometries (\ref{cmet}), (\ref{cg}) that 
\begin{equation}
f\frac{d\psi }{dx}+\frac{df}{dx}=\lambda (\frac{d\psi }{dy}+\frac{d\ln f}{dy}%
)\equiv A=\gamma -\delta \text{.}  \label{A}
\end{equation}
Then, the expression (\ref{tpl}) for $T_{1}^{1(PL)}$ can be rewritten, as 
\begin{equation}
T_{1}^{1(PL)}=-\frac{N}{96\pi f}[A^{2}-f_{x}^{\prime 2}]=-\frac{N}{96\pi f}%
[A^{2}-\lambda ^{2}\left( \frac{\partial \ln f}{\partial y}\right) ^{2}],
\label{t11}
\end{equation}
where we have taken into account the relation $\delta =-2\lambda $ ($\lambda 
$ determines the amplitude of the potential $U$ - see (\ref{U}), (\ref{ul})).

If at the right infinity the spacetime approaches the Minkowski form, the
parameter $B$ can be easily related to the effective temperature measured at
infinity \cite{thr}. Comparing (\ref{tpl}), (\ref{ps}) and (\ref{sol}), one
infers that 
\begin{equation}
\gamma =2\lambda (\frac{T}{T_{\lambda }}-1),B=\kappa (1-\frac{T^{2}}{%
T_{\lambda }^{2}}),A=4\pi T,  \label{cons}
\end{equation}
where asymptotically $T_{\mu }^{\nu (PL)}=\frac{\pi ^{2}NT}{6}^{2}diag(1,-1)$
and $T_{\lambda }=\frac{\lambda }{2\pi }$. It is worth noting that in the
particular case $T=0$ we get $B=\kappa $ and our solution (\ref{+-}) turns
into eq. (4.1) of \cite{cruz}.

Then it is convenient to rewrite (\ref{t11}) as 
\begin{equation}
T_{1}^{1(PL)}=-\frac{N}{6\pi f}[T^{2}-T_{H}^{2}]\text{,}
\end{equation}
where the Hawking temperature can be calculated in terms of the geometry
according to the standard rule 
\begin{equation}
T_{H}=\frac{k}{2\pi }=\frac{1}{4\pi }\left( \frac{df}{dx}\right)
_{x=x_{h}}=\lim_{y\rightarrow -\infty }\frac{\lambda }{4\pi }\frac{d\ln f}{dy%
}\text{,}  \label{th}
\end{equation}
$k$ is the surface gravity.

If, for black-hole solutions, one imposes the condition of finiteness of
quantum stresses in the frame of a free-falling observer on the horizon,
this condition singles out the unique value of temperature: $T=T_{H}$ \cite
{and}. The above condition is ensured by the choice $\gamma =0$, $%
T=T_{\lambda }$. In this case the function $\psi $ (\ref{ps}) is finite on
the horizon. One can also observe that, according to (\ref{sol}), $B=0$ as
well. Then (\ref{th}), (\ref{sol}) give us that for the exactly solvable
models under discussion $T_{H}=T_{\lambda }$ in accordance with \cite{exact}.

Let now $T\neq T_{\lambda }$, $\gamma \neq 0$, $B\neq 0$. In such a
situation we gain a free parameter $B$ (or $\gamma $), and, allowing it to
change, we may obtain $T\neq T_{H}$. Then, the existence of regular Killing
horizons becomes highly nontrivial issue for any $\kappa \neq 0$, however
small it be since according to (\ref{cons}), it leads to nonzero
coefficients $\gamma $, $B$, responsible for divergent stresses. The
situation can be also interpreted by saying that the true ''zero state'' of
the theory represents not a pure classical one but, rather, it incorporates
some essential quantum terms from the very beginning.

What would happen, if the contributions from higher loops were taken into
account, is not obvious in advance since their effect can be
model-dependent. Anyway, this does not mean that higher-loop effects would
necessarily destroy the character of the solutions under discussion. One may
speculate that, even if for some model accounting for higher loops does
destroy the phenomenon under discussion, it would be possible to insert the
corresponding higher corrections into the action coefficients $F$, $U$, $V$
from the very beginning (as it was with terms linear in $\kappa $ for RST or
CN\ model (\ref{cn})) and, repeating the same procedure in the higher
corrections, retain compatibility of regular geometries with infinite
quantum backreaction.

From the viewpoint of black hole physics, it is especially interesting that
the case $T=0$, when $T_{\mu }^{\nu (PL)}\rightarrow 0$ at infinity, $\,$%
also falls into the class of the solutions under consideration. It
represents the Boulware vacuum (vacuum with respect to the
Schwarzschild-like time at infinity). It is common belief that this state is
opposed to the Hartle-Hawking one in the following sense. In the Boulware
state the contribution of quantum stresses tends to zero at infinity but
cannot support a regular event horizon since it blows up there. In the
Hartle-Hawking one this contribution is finite and a regular horizon exists,
but quantum stresses at infinity tend to finite values representing thermal
radiation (unless the black hole is enclosed in a box). In our case,
however, we see that simultaneously (i) quantum stresses blow up on the
horizon, (ii) a regular horizon exists, (iii) if $T=0$, the contribution of
quantum stresses vanishes at infinity.

To summarize the contents of this subsection in few words, the basic idea is
the following. Usually the equality $T=T_{H}$ enables one to smooth out the
geometry. However, in the cases under consideration there is no need to
smooth it out since the physical (Lorentzian) geometry is already smooth
from the very beginning even in spite of $T\neq T_{H}$.

\subsection{Behavior of the action coefficients: weak-coupling regime near
the horizon}

On the first glance, one could try to ascribe unusual properties of the
solution under discussion to a singular behavior of the coupling coefficient
between curvature and dilaton since, indeed, $F$ diverges on the horizon for
the solutions. However, in this respect, it is important to note that the
role of the coupling effective ''constant'' between dilaton and curvature $%
g_{eff}$ is played not by $F$ itself, but by $\kappa F^{-1}\equiv
g_{eff}^{cl.}$ in the classical case or by $\kappa \tilde{F}^{-1}\equiv
g_{eff}^{q}$ in the semiclassical one, with quantum terms taken into
account. Substituting the explicit expression from (\ref{cn}) and (\ref{eq})
we get: 
\begin{equation}
g_{eff}^{cl.}=\kappa \exp (2\phi )\text{, }g_{eff}^{q}=\kappa [\exp (-2\phi
)+2\kappa d\phi ]^{-1}.  \label{gcl}
\end{equation}
Therefore, in the limit $\phi \rightarrow \infty $, which may correspond
just to the combination of regular geometry and divergencies of
Polyakov-Liouville stresses, as explained above, 
\begin{equation}
g_{eff}^{cl.}\rightarrow \infty \text{, }g_{eff}^{q}\backsim \frac{1}{\phi }%
\rightarrow 0\text{.}  \label{g0}
\end{equation}

Thus, for a pure {\it classical }system we would have the {\it strong}%
-coupling regime, but for the {\it quantum-corrected }one the phenomenon
under discussion occurs in he {\it weak}-coupling one. Thus, in the
near-horizon region we have rather an asymptotically free (similar to what
happens at Minkowski infinity), than singular behavior for this quantity.
Therefore, the conclusion about regularity of the geometry with infinite
stresses does not exceeds the bounds of validity of the semiclassical
approach.

It is instructive to write down the asymptotic behavior of all action
coefficients. For the solutions with infinite stresses but regular
geometries we have on the horizon ($y\rightarrow -\infty $, $\phi
\rightarrow \infty $): $F\sim \phi \sim -y\sim \tilde{F}\rightarrow \infty $%
, $U\rightarrow 0$, $V\left( \nabla \phi \right) ^{2}\sim f^{-1}\sim \exp
[2y(\rho -1)]\rightarrow \infty $ (recall, that $\rho <1$ for the solutions
at hand). On the other hand, in the region $y\rightarrow \infty $ of linear
dilaton vacuum, where the metric approaches the Minkowski form, the dilaton $%
\phi \rightarrow -\infty $, and we have for the CN model (\ref{cn}): $F\sim
e^{2y}\rightarrow \infty $, $U\sim e^{2y}\rightarrow \infty $, $V\left(
\nabla \phi \right) ^{2}\sim e^{2y}\rightarrow \infty $. We see that on the
horizon the divergencies of the action coefficients are even milder than at
infinity. Thus, the fact that all or some of action coefficients tend to
infinity indicates pathological features neither in the model Lagrangian nor
in the solutions themselves. Moreover, it is to the point to recall that,
when $B=0$, there is usually the domain of the strong coupling $g_{eff}\sim
1 $ near a horizon, whereas in our case $g_{eff}\rightarrow 0.$

\subsection{Exceptional case: finite stresses on the horizon despite $B\neq
0 $}

For our solutions (\ref{hor}), using (\ref{th}) and (\ref{cons}), we have 
\begin{equation}
T_{H}=T_{\lambda }(1-\rho )\text{, }\rho =\frac{1-z^{2}}{2d}\text{, }z\equiv 
\frac{T}{T_{\lambda }}\text{.}  \label{z}
\end{equation}
If we want to have from the left ($y\rightarrow -\infty )$ a regular
horizon, the condition (\ref{cond}) should be fulfilled. Let us pose the
following question. Is it possible to achieve $T=T_{H}$, with (\ref{cond})
satisfied, for $B\neq 0$? After some algebra, we obtain that this condition
leads to 
\begin{equation}
2d=1+z\text{, }z\leq \frac{1}{2}\text{.}  \label{dz}
\end{equation}
Thus, the answer is positive. It means that, if $1/2<d\leq \frac{3}{4}$, the
solution under discussion represents a black hole with an ''usual'' regular
horizon, on which quantum stresses remain finite. In this respect, it is
similar to black holes found in \cite{exact} but generalizes the
corresponding family (for which $\gamma =0=B$) to the case $\gamma ,B\neq 0$%
. For the exactly solvable models considered in \cite{exact} the Hawking
temperature $T_{H}=T_{\lambda }$ is determined solely by the amplitude $%
\lambda $ of the potential $U(\phi )$ (this fact was observed earlier for
CGHS and RST black holes \cite{callan}, \cite{rst}, \cite{solod}. Meanwhile,
now the Hawking temperature, according to (\ref{z}) and (\ref{dz}), equals $%
T_{H}=T_{\lambda }(2d-1)$ $=T$.

It is worth noting that, if $B\neq 0$, the action coefficients have near the
horizon the common asymptotic form for both cases - with either finite or
infinite stresses on the regular horizon since this behavior is determined
by the same eq. (\ref{sol}). This confirms one more time that nothing
pathological occurs with our model and all kinds of solutions should be
taken as ''equal in rights'' members of the same family.

\section{Extremal horizons}

\subsection{Explicit solutions and geometry}

In the previous work \cite{ext} it was observed that regular extreme
black-hole horizons can be consistent with infinite quantum backreaction.
However, this property was found for models with logarithmic dependence of
the curvature-coupling parameter on $\phi $ near the horizon. Such models
look not very realistic from the viewpoint of string theory. Below, we show
that the aforementioned property can be obtained for more realistic,
string-inspired models with the combination of exponents of $\phi $, if one
considers, instead of a generic exactly-solvable model (\ref{sol}), its
degenerate case that can be obtained by some limiting transitions.

Namely, let $\delta \rightarrow 0$ but $D\rightarrow \infty $ or $-\infty $
in such a way that the product $D\delta ^{2}=\Lambda $ remains finite. To
make it well-defined, one can write 
\begin{equation}
D=D_{0}+\frac{D_{1}}{\delta }+\frac{D_{2}}{\delta ^{2}}\text{,}
\end{equation}
where $D_{2}=\Lambda $ according to (\ref{const}). Then, after some
rearrangement we obtain the equation 
\begin{equation}
\tilde{F}^{(0)}=C^{\prime }-\frac{\kappa \gamma ^{2}}{4}\sigma ^{2}\text{,}
\label{f2}
\end{equation}
where $C^{\prime }$ is a new constant. According to (\ref{fd}), now 
\begin{equation}
f=e^{-\psi _{0}}=e^{2\phi }\text{,}
\end{equation}
where we choose, as usual, $\omega =-2$. Now the potential for our model
equals, according to (\ref{U}), (\ref{uf}) 
\begin{equation}
U=\Lambda e^{-2\phi }
\end{equation}
with $\Lambda =-\frac{\kappa \gamma ^{2}}{2}$. Let us take 
\begin{equation}
\tilde{F}^{(0)}=e^{\phi }-\kappa de^{-2\phi }  \label{new}
\end{equation}
with $d>0$. Then the dilaton field is an even function of $\sigma $, and it
follows from (\ref{f2}), (\ref{new}) that at $\sigma \rightarrow \pm \infty $%
, $f\sim \sigma ^{-2}\sim (x-x_{h})^{2}$, where $x$ is the Schwarzschild
coordinate. Such a quadratic dependence on the coordinate is just behavior
typical of the extremal horizons. As for the model (\ref{new}) $\tilde{F}%
^{(0)^{\prime }}>0$ everywhere including $\sigma \rightarrow \pm \infty $,
it is easy to check that the Riemann curvature remains finite and the
solution is everywhere regular. It is essential that, again, in this case $%
\gamma \neq 0$ and, for this reason, in (\ref{t11}) the constant $A=4\pi
T\neq 0$, $T_{H}=0$. Therefore, near the horizon $T_{1}^{1(PL)}$ diverges as 
$f^{-1}\,$but the geometry remains regular. It is worth noting that the
features under discussion are due to the quantum term (proportional to $%
\kappa $) and do not exist in the classical case ($\kappa =0$)$.$ In
particular, our model does not match smoothly the classical extremal black
hole recently considered in \cite{gukov}.

\subsection{Entropy of extremal horizons}

Traditionally, it was believed that black holes possess the
Bekenstein-Hawking entropy $S=A/4$ ($A$ is the area of the event horizon),
so it would seem that the limit from the nonextreme state to the extreme one
can be done directly. Actually, the thermodynamic behavior of near-extreme
black holes should be considered with great care due to the essential role
of quantum fluctuations \cite{mpl} and qualitatively new features in the
behavior of the entropy that in the extremal limit can tend to zero for
dilaton black holes \cite{wil}. Moreover, it was realized some time ago,
that thermodynamics of extreme black holes (EBH) can be qualitatively
different from that of nonextreme ones due to essentially different
topological properties in the Euclidean sector, and it was suggested to
ascribe arbitrary nonzero temperature and zero entropy to them \cite{ross} - 
\cite{gib}. The point, however, is that this prescription usually works only
in the tree (zero-loop) approximation. Quantum backreaction of fields,
surrounding a black hole, leads to divergencies of the stress-energy tensor
(SET) on the event horizon that destroy it completely \cite{and}, unless the
temperature is put to its Hawking value. As this value is zero for EBH, the
possibility of their thermodynamic description becomes questionable. It is
unclear whether the notion of entropy is applicable to such objects at all
and, if so, what is the value of the entropy. Recently, the possibility of a
thermodynamic description of EBH even without taking into account quantum
backreaction was placed in doubt in \cite{lib}, motivated\ by studying
dynamic process with ''incipient'' EBH - collapsing spherical bodies with an
exterior extreme Reissner-Nordstr\"{o}m metric. On the other hand,
calculations in string theory gave a definite value for the black hole
entropy of EBH but this value is the Bekenstein-Hawking one, so the property 
$S=0$ for the EBH black hole was not confirmed \cite{vaf}. Accounting for
quantum properties of spacetime makes the picture even more contradictory.
In particular, there are some arguments in the favour of the fact that the
wave function of EBH could vanish, thus forbidding the existence of EBH in
quantum theory \cite{claus}. On the other hand, on the semiclassical level
strong arguments in favor of existence of EBH were put forward in \cite{zero}%
, \cite{oj}, where it was observed that quantum backreaction may preserve
the extreme character of a horizon of the quantum-corrected
Reissner-Nordstr\"{o}m EBH.

In the absence of a full theory of quantum gravity it looks natural to
investigate carefully different possibilities that the semiclassical theory
supplies us with. As is said above, on the semiclassical level quantum
backreaction seems to invalidate the thermodynamic prescription made in \cite
{ross} - \cite{gib}. However, this argument does not apply to solutions of
the type discussed in the present article. Thermodynamic description of
these solutions and the value of the entropy should follow from the
Euclidean action formalism. As we shall see, the Euclidean action for such
solutions contains contributions from the horizon that differ from the
''usual'' case of classical extreme black holes and should be carefully
evaluated. We will see that such evaluation shows that for an extreme
solution of the type under discussion the Euclidean action is finite at
arbitrary nonzero temperature. Naive calculations give rise to non-zero
quantum corrections for the entropy but a more thorough treatment forces us
to introduce an additional inner boundary that cancels these terms and
confirms the property $S=0$ for the entropy of EBH.

One reservation is in order. Recently, there have appeared works on models
of two-dimensional (2D) dilaton gravity with non-minimal scalar fields \cite
{bur}, \cite{med} in which it was claimed that the SET of quantum fields in
the EBH background can be regular at arbitrary temperature, preserves the
regularity of the quantum-corrected geometry of EBH and is compatible with
the property $S=0$. However, the SET of quantum fields in \cite{bur}, \cite
{med} contains neither the temperature parameter nor other free parameters
explicitly, so it is rather difficult to check the claim made. Apart from
this, the derivation of the action for such models, as consistent and
reliable as for minimal fields, is still lacking, so the problems connected
with the extreme state overlap with problems inherent to 2D dilaton gravity
itself. All this deserves separate treatment but in the present paper we
restrict ourselves to the minimal fields for which the action describing
quantum corrections is well-defined (the Polyakov-Liouville action), there
exist explicit formulas for the SET in terms of the metric and it is certain
that SET at nonzero temperature cannot be regular on the extreme horizon 
\cite{and}.

By assumption, we consider spacetimes with Killing horizons. To elucidate
whether or not direct thermodynamic meaning can be assigned to them, we
should calculate the Euclidean action and check that it is finite.

Our main goal is to describe the properties of the extremal horizons and we
will only briefly discuss the nonextreme case. Then, our Euclidean manifold
has the topology of a disk and, if $T=\beta ^{-1}\neq T_{H}$, possesses a
conical singularity$.$ Taking account of such singularities is essential for
the calculation of the action and examining thermodynamic properties of the
system \cite{fis}. The Riemann curvature of the Euclidean manifold acquires
a conical singularity at the horizon (more exactly, the bolt that replaces
now the horizon of the Euclidean geometry) that is removed at the final
stage of calculations, when one puts $T=T_{H}$. Now this singularity
persists since, by definition, we consider just the solutions with $T\neq
T_{H}$. Meanwhile, a much more severe ''singularity'' reveals itself in
calculations than a pure geometrical conical one. Usually, the calculation
of the Euclidean action contains a contribution proportional to the value of
the coefficient $F$ on the horizon ($\tilde{F}$, if quantum correction are
taken into account) and responsible for the entropy. However, this
contribution is divergent and so is the total Euclidean action. This
confirms the observation, made in \cite{nonext} that thermodynamic
interpretation cannot be assigned to such horizons. Meanwhile, whatever
interpretation be suggested for the parameter $T\neq T_{H}$, the observation
that, in spite of infinite quantum stresses on the horizon, geometry is
regular there, retains its validity.

\subsection{Self-consistency of the variational procedure}

It is the issue of the thermodynamic of quantum-corrected extremal horizon
that we now turn to. Let us write down the metric in the form 
\begin{equation}
ds^{2}=a^{2}d\tau ^{2}+b^{2}dz^{2}\text{,}  \label{ab}
\end{equation}
where $0\leq z\leq 1$, $z=0$ corresponds to the horizon ($a(0)=0$) and $z=1$
corresponds to the boundary. In the Euclidean sector the action has the form 
\cite{action} 
\begin{equation}
I=-\frac{1}{2\pi }\int_{M}d^{2}x\sqrt{g}[R\tilde{F}(\phi )+\tilde{V}(\phi
)(\nabla \phi )^{2}+U(\phi )]+\frac{1}{\pi }\int_{\partial M}dsk\tilde{F}%
\text{,}  \label{eac}
\end{equation}
where $k$ is the second fundamental form, $ds$ is the line element along the
boundary $\partial M$ of the manifold $M$, and the Euclidean time $0\leq
\tau \leq \beta =T^{-1}$. If $n^{\mu }\,$is an outward vector normal to the
boundary, $k=-\nabla _{\mu }n^{\mu }$. It is convenient to normalize the
Euclidean time according to $\beta _{0}=2\pi $. The appearance of the titled
coefficients in the action is explained in Sec. II.

The variation of the action with respect to $\beta =2\pi a$ is expected to
have the general form 
\begin{equation}
\delta I=\int_{0}^{1}dz\tilde{T}_{0}^{0}b\delta \beta (z)+A_{1}\left( \delta
\beta \right) _{B}\text{,}  \label{exp}
\end{equation}
where $A_{1}$ is some coefficient. Then, if we fix the local inverse
temperature $\beta $ on the boundary, $(\delta \beta )_{B}=0$, we derive
from the action principle $\delta I=0$ the Hamiltonian constraint $\tilde{T}%
_{0}^{0}=0$ (the $00$ equation of the set of field equations).

Let the boundary consist of one point $B$, so integration in the action is
performed between a horizon and $B$. We will see below that, although for
''usual'' extreme black hole topologies eq. (\ref{exp}) holds, for our types
of solution we get instead 
\begin{equation}
\delta I=\int_{0}^{1}dz\tilde{T}_{0}^{0}b\delta \beta (z)+A_{1}\left( \delta
\beta \right) _{B}+A_{2}\left( \delta \beta \right) _{H}+A_{3}\delta \left( 
\frac{\partial \beta }{\partial l}\right) _{H}\text{,}  \label{dmod}
\end{equation}
where terms with $A_{2}$, $A_{3}$ in general do not vanish, the indices
''B'' and ''H'' refer to a boundary and horizon, respectively. Their
presence would spoil the variational procedure which implies that only the
boundary value of $\beta $ but not its value and normal derivative on a
horizon should be fixed. Then, the only way to escape this contradiction is
to introduce an additional fictitious boundary at $z=+0$. In other words,
the term $\frac{1}{\pi }\int dsk\tilde{F}$ in (\ref{eac}) should consist of
two parts and include not only the contribution from the physical boundary,
but also from the additional one. As a result, the terms stemming from a
horizon are killed since the horizon is surrounded now by a fictitious shell
and only the terms on the two pieces of the boundary may now contribute to $%
\delta I$. In other words, the terms with $A_{2}$, $A_{3}$ disappear but the
term with $A_{1}$ will include contributions from both pieces of the
boundary. This procedure leads to a self-consistent variational procedure
and (with some restriction on the behavior of the action coefficients which
should not grow near the horizon too rapidly) to a finite Euclidean action,
from which one finds the value of the energy and entropy (see below).

Direct calculation gives us, after simple rearrangement, that the original
action (\ref{eac}) can be written down as 
\begin{equation}
I=\int_{0}^{1}dz\tilde{T}_{0}^{0}\beta b+I_{1}\text{,}  \label{niz}
\end{equation}
\begin{equation}
2\pi \tilde{T}_{0}^{0}=2\frac{\partial ^{2}\tilde{F}}{\partial l^{2}}-\tilde{%
V}\left( \frac{\partial \phi }{\partial l}\right) ^{2}-U\text{.}  \label{hc}
\end{equation}
Here $I_{1}$ stems from the term outside the integral after integration by
parts plus the boundary term: $I_{1}=I_{2}+I_{3}$, where 
\begin{equation}
I_{2}=-\frac{1}{\pi }\left( \tilde{F}\frac{\partial \beta }{\partial l}%
\right) _{H}\text{,}  \label{ni2}
\end{equation}
\begin{equation}
I_{3}=\frac{1}{\pi }\left( \beta \frac{\partial \tilde{F}}{\partial l}%
\right) _{H}-\frac{1}{\pi }\left( \beta \frac{\partial \tilde{F}}{\partial l}%
\right) _{B}\text{,}  \label{ni3}
\end{equation}
and $dl$ is the proper distance element. For the extreme case there are no
conical singularities, and the topology corresponds to the annulus, whose
inner boundary lies at an infinite proper distance \cite{teit}.

\subsection{Classical EBH}

For ''usual'' EBH $\tilde{F}$ is finite on the horizon. Take now into
account that the Hawking temperature $T_{H}=\frac{1}{2\pi }\left( \frac{%
\partial a}{\partial l}\right) _{H}=0$ for extreme black holes. Then we see
that the term $I_{2}$ vanishes. As $a=0=\beta $ on the horizon, in the term $%
I_{3}$ the horizon contribution vanishes and only the boundary one survives.
As a result, we get $I=\int_{0}^{1}dz\tilde{T}_{0}^{0}\beta b-\beta \frac{1}{%
\pi }(\frac{\partial \tilde{F}}{\partial l})_{B}$. We take into account
that, according to (\ref{hc}), the quantity $\tilde{T}_{0}^{0}$ does not
contain $\beta $. Then, the variation with respect to $\beta $ takes the
general form (\ref{exp}). Inserting the Hamiltonian constraint $\tilde{T}%
_{0}^{0}=0$ into the action, we obtain 
\begin{equation}
I_{tot}=\beta _{B}E\text{,}  \label{ecl}
\end{equation}
\begin{equation}
\beta _{B}=2\pi a_{B}\text{, }E=-\frac{1}{\pi }(\frac{\partial \tilde{F}}{%
\partial l})_{B}\text{,}
\end{equation}
''B'' refers to the boundary, the quantity $E$ has the physical meaning of
the energy, the entropy $S=0$ in accordance with the conclusion of \cite
{ross} - \cite{gib}\ (see also \cite{kum} for 2D dilaton black holes).

\subsection{Modification of Euclidean approach for self-consistent extreme
solutions at $T\neq 0$}

This is just the main point of our consideration of the issue of entropy.
The quantity $(2\pi )^{-1}\frac{\partial \beta }{\partial l}\rightarrow
\kappa $ on the horizon, where $\kappa $ is the surface gravity. On one
hand, $\kappa =0$ since the geometry, by assumption, corresponds to the
extreme case. But, from the other hand, the quantity $\tilde{F}\rightarrow
\infty $. Thus, we have the undetermined product of two competing factors.
As a result, the quantity $I_{2}$ would not in general vanish and, thus, it
would contribute to (\ref{dmod}) (the term with $A_{3})$ that, as is
explained above, would spoil consistency of the variational procedure. In (%
\ref{ni3}) the first term also turns out to be the product of zero and
infinite quantities and generates the term with $A_{2}$ in (\ref{dmod}). We
already know what to do: it is necessary to kill such terms due to
introducing an additional boundary at $z=+0$. Then, direct calculation of
the Euclidean action (\ref{eac}), where now the boundary term includes
contributions not only from the physical boundary at $z=1$, but also from
the fake one placed on the horizon, gives us 
\begin{equation}
I=\beta _{out}E_{out}+\beta _{in}E_{in}\text{,}  \label{i2}
\end{equation}
where 
\begin{equation}
\beta _{out}=2\pi a(1)\text{, }\beta _{in}=2\pi a(+0)\text{,}
\end{equation}
\begin{equation}
E_{out}=-\frac{1}{\pi }(\frac{\partial \tilde{F}}{\partial l})_{z=1}\text{, }%
E_{in}=\frac{1}{\pi }\lim_{z\rightarrow 0}(\frac{\partial \tilde{F}}{%
\partial l})\text{.}  \label{e2}
\end{equation}

Comparing with the general thermodynamic form of the action for a system
with a boundary, consisting of two pieces (two shells in thermal
equilibrium), $I=\beta _{out}E_{out}+\beta _{in}E_{in}-S$, we conclude that
the entropy $S=0$. In principle, as the quantity $\tilde{F}$ enters these
products, the properties of the system are model-dependent. Moreover, in the
action $\beta _{in}\rightarrow 0$, $E_{in}\rightarrow \infty $, so the form (%
\ref{i2}) does not guarantee the finiteness of the action if $\tilde{F}$
diverges near the horizon (inner boundary) too rapidly. Let us restrict
ourselves by a general non-degenerate case (\ref{eq}), (\ref{cg}). Then,
simple evaluation, exploiting the explicit form of the solutions (\ref{sol}%
), shows that the product $\beta _{in}E_{in}$ remains finite.

It is worth stressing that our scheme for calculating the action and entropy
is quite general, so the condition of exact solvability may be relaxed in
this respect. The expression for the action can be rewritten in the
conformal gauge in the form 
\begin{equation}
I=-2\lambda (\frac{\partial \tilde{F}}{\partial y})_{z=1}+2\lambda (\frac{%
\partial \tilde{F}}{\partial y})_{z=0}\text{.}
\end{equation}
Therefore, to obtain a finite action, one only needs that the coefficient $%
\tilde{F}$ grows near the horizon not more rapidly than the first degree of $%
y$.

Let us summarize the basic steps that led us to the final result about the
entropy. The divergencies in the action coefficient $F$ result in the
failure of the standard variational procedure due to terms stemming from a
horizon. The only way to repair it is to introduce an additional boundary
before a horizon that automatically excludes the potential entropy
contribution and gives us the value $S=0$. Thus, the divergencies which
usually manifest themselves as a stumbling-block in attempt to expand the
notion of the entropy from classical extremal horizon to quantum-corrected
extremal ones (because of thermal divergencies caused by the inequality $%
T\neq T_{H}=0$), now themselves suggest how to solve the problem and give a
quite definite answer.

It is worth noting that the inner boundary for extremal horizons of 4D
dilaton black holes was suggested in \cite{gib} with the aim of obtaining
the integer value for the Euler characteristics and an unambiguous answer
for the entropy (cf. also discussion of the role of the horizon in black
hole thermodynamics of nonextreme and extreme black holes in \cite{teit}).
There are, however, two essential differences between the situations
discussed in \cite{ross} - \cite{gib} and the present one. First, black
holes, considered in the aforementioned articles, were purely classical, the
corresponding approach being applied to the case when the
gravitation-dilaton coupling is finite on the horizon (for example, in the
case of general relativity, $F=1$), whereas in our case quantum backreaction
is crucial and it has divergencies in the coefficient $\tilde{F}$ due to
this backreaction that enforced us to introduce the additional inner
boundary. Second, the approach elaborated in \cite{ross} - \cite{gib} shows
the difference between the thermodynamics of {\it classical nonextreme} and 
{\it classical extreme }black holes, whereas our approach handles the
difference between {\it classical extreme }and {\it quantum-corrected extreme%
} ones.

\subsection{Discussion: peculiarities of the energy and entropy of extremal
horizons in the given context}

We see that the Euclidean action we dealt with turned out to be finite, with
the black hole entropy $S=0$. The price we paid for it is the divergencies
in the energy associated with the horizon. This features looks quite unusual
but, in our view, nothing unphysical appears here. To clarify this point,
let me refer to the following analogy. In the reduction procedure from 4D
spherically-symmetrical theories to 2D ones the effective dilaton field is
introduced through the radial coordinate according to $r=\exp (-\phi )$. If $%
\phi \rightarrow \infty $ , $r\rightarrow 0$. In this sense, the analogy
between this limit and the point $r=0$ of 4D spacetime can be carried out 
\cite{flux}. However, in the quantum corrected case (the situation we deal
with) the effective $r^{2}$ (the quantity similar to our $\tilde{F}$)
acquires terms growing as $B\left| y\right| $ (where the term $B$ has a pure
quantum origin) near the horizon and, thus, diverges. In terms of the
corresponding 4D theory, this would mean a black hole with an infinite area
of an event horizon. Fortunately, such objects have already been found in 4D
gravity - mainly, in Brans-Dicke theory \cite{jor} - \cite{cold} and are
shown to be well-defined, with curvature invariants bounded on the horizon
(at least, for some sets of parameters). Moreover, a recent study showed
that their thermodynamics is also well-defined and it was shown \cite{inf}
that the effective quasi-local energy density (per unit area) turns out to
be finite, but the total energy diverges because of the infinite area of the
horizon. In our 2D case an infinite $\tilde{F}$ can be thought of as
reminiscent of an infinite area in 2D theory and, thus, the divergencies in $%
E_{in}$ look quite natural (but, let me stress it again, the total action is
finite).

It is also instructive to note that accounting for quantum backreaction does
not change the relationship between the entropy and the Euler
characteristics $\chi $. Indeed, under the shift $\psi \rightarrow \psi +C$
the Polyakov-Liouville action changes according to $I_{PL}\rightarrow
I_{PL}+2\kappa \chi C$, where $\chi =1$ for the nonextreme case \cite{solcon}%
, \cite{solod} and $\chi =0$ for the extreme one. Let us denote the
contribution of thermal gas situated between the boundary, enclosing a
system, and a horizon, as $S_{q}$. Then, in the first case, it is natural to
fix the constant by the demand that $S_{q}\rightarrow 0$ when $x\rightarrow
x_{H}$ (no room for quantum radiation). This condition loses its sense for
EBH since the proper distance between the horizon and any other point is
infinite. In fact, one does not need to impose such a condition at all since
the action and the value of the entropy $S=0$ are not influenced by the
choice of $C$ due to the factor $\chi =0$.

Let us also to summarize the results and enumerate briefly some distinct
features of thermodynamics of extremal horizons in the given context. (1)
For nonextreme black holes the total entropy $S=S_{bh}+S_{q}$, where $S_{bh}$
is the Bekenstein-Hawking entropy or its two-dimensional analogue. However,
now we obtained $S=0$ for the {\it total} entropy, and there is no separate
analogue of $S_{q}$ in spite of the fact that temperature is nonzero. (2)
For extreme but classical black holes it is obvious in advance that the
Euclidean action is finite. Now it was not so obvious because of the
infinite behavior of the action coefficients at the horizon. Nonetheless,
the final answer is indeed finite (under some, not very severe, restrictions
on the behavior of the coefficient $\tilde{F}$ near the horizon). (3) If the
Euclidean action is taken in the standard Hilbert form, the variation
procedure fails to be self-consistent. This is repaired by introducing an
additional boundary that moves in the direction of the horizon. The
corresponding energy, associated with this boundary, diverges, although the
action itself is finite. This is the price, paid for a well-defined
Euclidean action in the given context.

\section{Relevance for 4D world: black holes without thermodynamics?}

In the preceding sections, we showed, using {\it exactly }solvable models,
that semiclassical nonextreme black holes (with one-loop quantum
backreaction taken into account) without the property $T=T_{H}$ can, indeed,
exist in two-dimensional (2D) dilaton gravity \cite{nonext}. Although one
can always calculate $T_{H}$, expressing its through the geometrical
characteristics of the horizon, such a quantity does not determine in the
aforementioned case the temperature of quantum fields. As a result, an
intimate link between quantum theory, properties of the horizon and
thermodynamics is broken, so thermodynamic interpretation cannot be assigned
to such exceptional black hole solutions. Apart from this, quite recently it
was demonstrated in \cite{inf} (using {\it exact} solutions found in \cite
{bbm}, \cite{bbm2}) that in the pure 4D dilaton gravity with conformal
coupling the Euclidean action diverges. This means that such black holes
cannot also be considered as thermodynamics objects.

The fact that some exact solutions both in 2D and 4D gravity exhibit such
unusual properties forces us to take this point seriously and to try to
understand better in which cases the exceptional black hole solutions of
this kind may arise. Now we are trying to combine both a more realistic (but
more complicated) 4D theory and quantum backreaction. We are unable,
obviously, to find exact solutions in such a situation, but as we will see,
the analysis of the behavior of a system near the horizon is quite
tractable. It is worth noting that we do not pretend to carry out analysis
for some concrete realistic models. Instead, we only elucidate under which
general conditions quantum backreaction and regular geometry near the
horizon can be consistent without the demand $T=T_{H}$. In our view, such an
approach is to be justified, when it is applied to the issue of such a
general character as fundamentals of black hole thermodynamics.

For dilaton theory, the classical part of the action reads 
\begin{equation}
I=\frac{1}{16\pi }\int d^{4}x\sqrt{-g}[RF(\phi )+V(\nabla \phi )^{2}+U(\phi
)]\text{.}  \label{action}
\end{equation}
Then the field equations take the form (\ref{feq}), (\ref{feq1}), where $%
T_{\mu }^{\nu (m)}$ is the average value of the quantum stress-energy tensor
(SET), describing backreaction of quantum fields on the geometry, $G_{\mu
}^{\nu }$ is the Einstein tensor. In general relativity $F=-1$, $%
U=V=0=T_{\mu }^{\nu ^{\phi }}$. Then it is obvious that divergent $T_{\mu
}^{\nu (m)}$ and finite $G_{\mu }^{\nu }$ are mutually inconsistent.
However, we will see below that in some models of dilaton theory this is
indeed possible due to mutual compensation of divergencies (which occur on
the horizon) of all contributions in (\ref{feq}).

Formally, one can always achieve the equality $F=-1$ by a suitable conformal
transformation. However, as we will be dealing with the situation when the
factor $F$ tends to zero or infinity, both spacetimes become physically
non-equivalent - for instance, one of them may be regular, whereas the
second one is not. Therefore, we will retain the general form of the action (%
\ref{action}).

Let us consider a spherically-symmetrical static spacetime. Its metric takes
the form 
\begin{equation}
ds^{2}=-fdt^{2}+f^{-1}dr^{2}+R^{2}d\Omega ^{2}\text{,}  \label{m}
\end{equation}
where $d\Omega ^{2}=d\theta ^{2}+\sin ^{2}\theta d\phi ^{2}$ and by proper
rescaling of the radial coordinate we achieved the equality $g_{00}g_{11}=-1$%
. In what follows we will assume that there exists a horizon at $r=r_{+}$.
In particular, for black holes in string theory \cite{gib1}, \cite{maeda}, 
\cite{4dil} $f=1-\frac{r_{+}}{r}$, $R^{2}=r(r-r_{0})$, where $r_{0}$ is
proportional to $Q^{2}$ ($Q$ is an electric charge). However, to avoid
unnecessary complication, not connected with the essence of matter, we also
assume that the electromagnetic field and corresponding charges are absent.
In the spherically-symmetric case we have only three independent equations: $%
00$, $rr$ and $\theta \theta $ ones.

The exact form of SET cannot be found in an explicit form and its
approximate expression is very cumbersome. Therefore, on first glance, the
task to find and analyze concrete types of self-consistent solutions with
quantum backreaction looks absolutely hopeless. Fortunately, what we need is
only the asymptotic form near the horizon. Let us consider, for
definiteness, scalar massless conformal fields. Then in the thermal state 
\cite{scalar} 
\begin{equation}
T_{\mu }^{\nu (m)}=\frac{A}{16\pi f^{2}}diag(-1,\frac{1}{3},\frac{1}{3},%
\frac{1}{3})+\left( T_{\mu }^{\nu }\right) _{reg}\text{.}  \label{split}
\end{equation}
Here $\left( T_{\mu }^{\nu }\right) _{reg}$ is the part of SET that remains
regular on the horizon, $A=\bar{A}(T^{4}-T_{H}^{4})$, where $T$ is
temperature measured by a distant observer, $T_{H}$ is the Hawking
temperature, and $\bar{A}=\frac{N\pi ^{2}}{480}$, where $N$ is a number of
fields. If $T=T_{H}$, $A=0$ and we obtain the Hartle-Hawking state in which
SET is finite on the horizon. (For nonconformal or massive fields the
divergent terms contain also contributions of the order $f^{-1}$, as follows
from eq. (4.4) of Ref. \cite{scalar}).

One reservation is in order. The expression for SET derived in \cite{scalar}
with using WKB approximation contains also logarithmic terms divergent on
the event horizon which persist even in the Hartle-Hawking state ($T=T_{H}$%
). However, they seem to be an artifact of the particular perturbative
scheme (that becomes not quite adequate near the horizon). For example, for
massive fields calculations based on the Schwinger - DeWitt approximation 
\cite{matrn} give no indication of such terms independent of the concrete
form of the static metric. The modified versions of WKB approximation also
show that there are no logarithmic terms for the mean values of $\phi ^{2}$
in the Hartle-Hawking state for a generic spherically-symmetric spacetime 
\cite{sergey} (see also the analysis of the Reissner-Nordstr\"{o}m
background in \cite{tom}). Numerical computations in \cite{scalar} also
testify against the logarithmically divergent terms. It would be tempting to
substantiate eq. (\ref{split}) (and its counterpart for massive fields)
without refereeing to explicit particular computational scheme, but for the
present such a rigorous proof is lacking.

We pose the question: is it possible to get a regular black hole geometry as
a solution of field equations in spite of divergencies on the horizon, where 
$f$ vanishes? We must compensate the leading divergencies in SET of quantum
fields. If we succeed with this, further terms of asymptotic expansion can
be found from Taylor series near the horizon, as corrections. We want to
adjust the action coefficients (i.e., fix the model)\ in such a way that
near the horizon 
\begin{equation}
F_{\mu }^{\nu }\equiv \nabla _{\mu }\nabla ^{\nu }F\simeq \gamma _{\mu
}^{\nu }f^{-2},\square F\simeq \gamma f^{-2}\text{, }U\simeq \beta f^{-2}%
\text{, }V(\nabla \phi )^{2}\simeq \alpha f^{-2}\text{,}  \label{exp}
\end{equation}
where $\alpha $, $\beta $, $\gamma $, $\gamma _{\nu }^{\mu }$ are some
constants. The term $FG_{\mu }^{\nu }$ has the main order $f^{-1}$ and does
not contribute to the leading divergencies. Equating all terms of the order $%
f^{-2}$, we obtain from (\ref{feq}), (\ref{feq1}) the linear system

\begin{equation}
2(\gamma -\gamma _{0}^{0})-\beta -\alpha =A\text{,}  \label{0}
\end{equation}
\begin{equation}
2(\gamma -\gamma _{1}^{1})-\beta +\alpha =-\frac{A}{3}\text{,}  \label{1}
\end{equation}
\begin{equation}
2(\gamma -\gamma _{2}^{2})-\beta -\alpha =-\frac{A}{3}\text{,}  \label{2}
\end{equation}
where $x^{0}=t$, $x^{1}=r$, $x^{2}=\theta $, $x^{3}=\phi $. First, let us
consider the nonextreme case, when $f\backsim r-r_{+}$ near the horizon. Let
us try to choose the asymptotic 
\begin{equation}
F\simeq F_{0}+F_{1}f^{-1}  \label{asf}
\end{equation}
to obtain the desired behavior $f^{-2}\backsim (r-r_{+})^{2}$ for $F_{\mu
}^{\nu }$. Simple calculations show that it is indeed compatible with (\ref
{exp}), provided $\gamma =F_{1}f^{\prime 2}(r_{+})$, $\gamma _{0}^{0}=-\frac{%
\gamma }{2}$, $\gamma _{1}^{1}=\frac{3}{2}\gamma $, $\gamma _{2}^{2}=0$,
where the prime denotes differentiation with respect to $r$. We took into
account that, as a black hole by assumption is nonextreme, $f^{\prime
}(r_{+})=4\pi T_{H}\neq 0$, the function $R^{2}$ and its derivatives are
finite and nonzero near the horizon. Substituting the explicit expression
for $\gamma _{\mu }^{\nu }$ into (\ref{0}) - (\ref{2}), we obtain the
solution: $\gamma =\frac{4}{3}A$, $\beta =A$, $\alpha =2A$. This guarantees
that the field equations (\ref{feq}), (\ref{feq1}) are fulfilled near the
horizon. We have also to express the action coefficients in terms of the
dilaton. In other words, we adjust our model to the asymptotics we need. Far
from the horizon the form of the action coefficients is not restricted.

Let the horizon correspond to $\phi \rightarrow \infty $ and let $f\simeq
f_{1}\phi ^{-1}$ near it, then 
\begin{equation}
V\simeq V^{(0)}+V_{1}\phi ^{-1}\text{, }F\simeq F_{0}+\frac{F_{1}}{f_{1}}%
\phi \text{, }U\simeq \frac{\beta }{f_{1}^{2}}\phi ^{2}\text{,}  \label{vne}
\end{equation}
where $f_{1}$ is one more constant, $V_{1}=\alpha f_{1}^{-1}f^{\prime
-2}(r_{+})]$, $V^{(0)}\ll V_{1}\phi ^{-1}$ near the horizon (for instance, $%
V^{(0)}=V_{0}e^{-2\phi }$).

Next, consider the extreme case, when $f\simeq f_{0}\frac{(r-r_{+})^{2}}{2}$
(to avoid possible confusion, recall that, in contrast to singular extremal
solutions for charged dilaton black holes \cite{4dil}, we are looking for
regular ones only, so the function $R(r)$ is regular near the horizon). Now,
the asymptotic form (\ref{asf}) is not suitable since it does not generate
the terms $f^{-2}$, necessary for compensation of those in SET. Let us try
to choose instead 
\begin{equation}
F\simeq F_{0}+F_{1}f^{-2}  \label{ase}
\end{equation}
near the horizon. In this case the first term in (\ref{feq}) should be taken
into account. Near the horizon $G_{\mu }^{\nu }=(-\frac{1}{r_{+}^{2}},-\frac{%
1}{r_{+}^{2}},f_{0},f_{0})$. We have three equations (\ref{0}) - (\ref{2})
for four quantities $\alpha $, $\beta $, $\gamma $, $f_{0}$ which can be
expressed in terms of $A$ and $r_{+}$. We may choose $\phi \simeq \frac{\phi
_{0}}{f^{2}}$ near the horizon and $U\backsim \phi $, $V\simeq V^{(0)}(\phi
)+V_{1}\phi ^{-1}$, where, again $V^{(0)}(\phi )$ decays faster than $\phi
^{-1}$.

One may wonder, whether or not the possibility to combine the regularity of
geometry with infinite quantum stresses on the horizon arises due to
singularity in the gravitation-dilaton coupling $g_{eff}$. However, the
analysis performed in Sec. IVD for the 2D case, applies here directly and
leads to the conclusion that, although classicaly $g_{eff}$ indeed diverges,
the semiclassical version of $g_{eff\text{ }}$ remains finite and even
vanishes. Thus, the effect under consideration occurs in the weak-coupling
regime, where semiclassical approximation can be trusted.

I would like to stress that the procedure I follow looks very much like the
usual quasi-classical scheme in that we take the SET evaluated on the given
background. There are two important differences, however: (i) we write SET
for a metric, which is unknown in advance, and solve the corresponding
system of field-equations {\it self-consistently}; more exactly, as it is
absolutely impossible to find the exact solutions in the whole region, I
consider only asymptotic behavior of the metric and dilaton near the
horizon; (ii) usually there exist classical solutions (both for a dilaton
and metric) to which quantum backreaction adds small corrections; here, by
contrast, the corresponding classical solutions lose their meaning in the
absence of quantum backreaction. It is also worth stressing that the
geometry obtained in this approach as a result of strong backreaction is
classical in the sense that the curvature scale is far from the Planck
regime.

As the solutions under discussion possess at once several unusual
properties, it would be nice to confirm them, using some explicit examples
of exact solutions. Unfortunately, because of the high complexity of
quantum-dilaton-gravitation equations in the 4D case, this is impossible.
However, the fact that 2D theories (see above) do possess {\it exact}
solutions with properties described above ($T\neq T_{H}$, but the curvature
on the horizon is finite) forces us to take the phenomenon seriously in the
4D world as well.

Thus, the general scheme consists of the following. To construct a dilaton
model, suitable for our purposes, we adjust at our will the action
coefficients $F(\phi )$, $U(\phi )$, $V(\phi )$ $\,$in such a way that near
the horizon they give a divergent contribution to the field equations to
compensate that from the quantum SET. This procedure can be interpreted as
quantum deformation of some original classical solutions since our
additional terms have near the horizon the same magnitude as the quantum
contributions from SET\ (but with the opposite sign). The quantum part
contains the terms with the coefficients that vanish in the classical limit
but, if they are non-zero, the main contribution near the horizon comes just
from them.

In fact, we perform the quantum deformation of the original action
coefficients, that can be compared with a similar procedure suggested in 
\cite{rst} for 2D dilaton gravity, where one adds ''by hand'' to the
classical gravitation-dilaton action some terms that contain the
quantum-coupling parameter. However, while in the case of 2D models the goal
was to make the model exactly solvable, now we want to ensure the existence
of solutions with regular horizon and infinite quantum backreaction, not
demanding exact solvability. Meanwhile, I want to stress that the RST\ model 
\cite{rst} contains, apart from other types of solutions, also those of the
considered type (with infinite backreaction on the horizon but regular
geometry) \cite{nonext}. It is worth noting that our consideration is purely
local and restricted to the region near the horizon. As far as a global
structure of spacetime is concerned, we can only point out that no- go
theorems for Einstein equations with scalar field (see recent papers \cite
{lem}, \cite{kir} and literature quoted there) cannot be used in our
situation for two reasons: (i) as the quantity $F\rightarrow \infty $ at the
horizon, the conformal transformation to the Einstein frame leads to a new
system, which is not equivalent to the original one, (ii) apart from the
scalar field (dilaton), the quantum matter source is present, both
contributions being divergent on the horizon but compensating each other. On
the other hand, the general logic on which our approach is based, is
applicable in more complicated situation, when, apart from the scalar field
or dilaton, other field (electromagnetic, Yang-Mills, etc.) are present. The
approach elaborated in the present paper can also be extended directly to
many-dimensional cases.

Thus, at least for some special models, black holes of the considered type
may exist in dilaton gravity. For these black holes the Hawking temperature
itself can be calculated according to the standard relation $T_{H}=\frac{%
\kappa }{2\pi }$ ($\kappa $ is the surface gravity) but it loses its
significance in this exceptional case. Indeed, the intimate link between
gravitation, spacetime and thermodynamic is broken in the sense that now we
are not obliged to put $T=T_{H}$ for quantum fields since the Lorentzian
geometry near the horizon is smooth from the very beginning, and there is no
need to make additional efforts to smooth it out. Because of separation of
geometry and thermodynamic properties, it would be very important to trace
whether black hole evaporation is still present for such solutions.

The essential feature of our solutions consists in that $F$ diverges on the
horizon. One may ask whether this can spoil a regular character of
spacetimes. In this regard, we would like to stress that, if the functions
have the asymptotics $f=f^{\prime }(r_{+})(r-r_{+})$, $R(r)=R(r_{+})+R^{%
\prime }(r_{+})(r-r_{+})$ (as was assumed above in the nonextreme case), it
is straightforward to show that the curvature tensor remains bounded not
only in the static frame, but in that of a free-falling observer as well.
The same is true for the extreme case if near the horizon $%
f=f_{0}(r-r_{+})^{2}$, $R(r)=R(r_{+})+R^{\prime }(r_{+})(r-r_{+})$. It seems
to the point to recall a known solution for a classical gravitating
conformal scalar field \cite{bbm} for which the quantity $F$ (if it is
rendered in our notations) also diverges on the horizon but this does not
cause any physical inconsistencies \cite{bek75}. This solution turned out to
be unstable against linear perturbations \cite{uns} that can be
qualitatively explained by vanishing $F$ at some point \cite{bek98}.
However, in our case we can adjust $F$ far from a horizon at our own will,
so it seems that the origin of this instability can be removed.

There is another potential origin of instability for the solutions under
discussion, connected with higher order quantum corrections. It may happen
that the answer to the question whether these corrections destroy the
character of our solutions, is model-dependent and cannot be done in an
universal form. On the other hand, it looks also quite probable that,
fine-tuning the coefficients of the gravitation-dilaton part of the action,
one can generate counterparts that kill dangerous terms coming from higher
orders in the same manner as it was done in the one-loop approximation (see
above). In our view, independent of whether or not the solutions under
discussion can be stable, they may be of interest in what concerns the
fundamentals of black hole thermodynamics. They point to some isolated gaps
in the standard picture which can exist as the manifestation of the
qualitative distinction between general relativity and dilaton (scalar)
gravity theories.

In other words, semiclassical theory of gravity (quantized matter fields
along with classical metric and dilaton) contains, if taken consistently,
quite unusual predictions within its own framework, and it was the aim of
our article to draw attention to the existence of such phenomena which are
not restricted by low-dimensional models.

\section{Summary and conclusion}

We examined a series of exactly solvable models of 2D dilaton gravities and
showed that the combination of regular geometry with infinite contribution
of quantum stresses looks quite typical of 2D\ dilaton gravity and should
not be considered as a rare exception. Several well-known exactly solvable
models share these properties which did not receive proper attention before.
The phenomenon under discussion concerns both nonextreme and extreme black
holes and occurs in the region of a weak effective gravitation-dilaton
coupling, where semiclassical approximation can be trusted.

The suggested types of solutions enabled us to find a {\it self-consistent}
closed thermodynamic interpretation of extremal Killing horizons that goes
beyond the tree level approximation and persists on the {\it semiclassical}
level too. In fact, {\it any} attempt to ascribe a definite value of the
entropy to extremal horizons should take into account the appearance of
infinite stresses on them due to deviation of temperature from its (zero)
Hawking value. We coped with this task in a general form, without appeal to
exact solvability. The only restrictions, necessary for the finiteness of
the Euclidean action, come from the demand that the action coefficient $%
\tilde{F}$ grow near the horizon as first degree of a $y$ (conformal
coordinate) or slower.

Thus, on one hand, thermodynamic interpretation in the semiclassical region
fails for nonextreme horizons but, on the other, is justified for extreme
ones - the usual picture is turned over.

Similar effects seem to exist in the 4D case, when exactly solvable
semiclassical models are absent but the idea remains the same: if a physical
Lorentzian geometry is smooth irrespective of the value of temperature,
there is no need to try to smooth it out by putting the temperature equal to
its Hawking value.

The essential feature of the models considered in the present paper consists
in that fluxes of dilaton and quantum matter fields become infinite on the
horizon each separately. If a device measuring each of them can be
constructed, it would probably mean that a horizon for an observer endowed
with such a distinctive detector would remain unattainable. Thus, one would
get a black hole with a regular horizon which, however, cannot be crossed by
any observer - to some extent, this can be considered as a quantum analogue
of naked black holes \cite{naked1}, \cite{naked2}. Let me recall also that
divergencies of quantum stresses (although more mild) inevitably occur for a
free falling observer in the metric of an extreme black hole, even if these
stresses remain bounded in the static frame \cite{triv}.

It remains unclear how the account of higher-order quantum corrections,
including those in the dilaton and the metric, can modify the picture
described in the present article. However, the very fact that the system,
governed by the action (\ref{tot}) - (\ref{pl}) and considered as
self-closed, may exhibit the behavior discussed above, deserves, in our
view, attention. Even if some models (especially, in 4D case) may look not
very realistic from the viewpoint of concrete applications, their advantage
consists in that they show that the phenomenon under discussion is possible
in principle.

The existence of regular geometries even despite divergencies of quantum
matter field stresses can also suggest some new approach to the problem of
singularities in the theory of gravitation and open new possibilities in
cosmological scenarios.

At the dawn of black hole thermodynamics, whose beginning was marked by
papers of Prof. Bekenstein \cite{bekenst}, it was a great surprise, which
can scarcely be exaggerated, that black holes possess their own
thermodynamic properties. Now, it is the importance of this phenomenon that
forces to draw special attention to the potential exceptions in this picture.

{\it Note added}. After submission of this paper, we have managed to extend
analysis beyond exactly solvable models and showed that infinite quantum
backreaction and regularity of a horizon may be compatible for systems with
coefficients $\tilde{F}$, $\tilde{V}$, finite on a horizon \cite{mod}.

\section{Acknowledgement}

I am grateful to Claus Kiefer and Sergei Solodukhin for encouraging interest
to this line of research, and to Paul Anderson and Jurek Matyjasek for
helpful correspondence. I\ am also grateful to Dmitri Kazakov for one
question on a seminar in Dubna many years ago.





%
%

%
%

\end{document}